# Modification of special relativity and the divergence problem in quantum field theory

Jian-Miin Liu*
Department of Physics, Nanjing University
Nanjing, The People's Republic of China
*On leave. E-mail address: liu@mail.davis.uri.edu

The root of the divergence problem in the current quantum field theory seems to be in the special theory of relativity. Here we propose a modified special relativity theory by introducing the primed inertial coordinate system, in addition to the usual inertial coordinate system, for each inertial frame of reference, assuming the flat structures of gravity-free space and time in the primed inertial coordinate system and their generalized Finslerian structures in the usual inertial coordinate system, and combining this assumption with the two fundamental postulates, (i) the principle of relativity and (ii) the constancy of the one-way speed of light in all inertial frames of reference. The modified special relativity theory involves two versions of the light speed, infinite speed c' in the primed inertial coordinate system and finite speed c in the usual inertial coordinate system. The physical principle is: the c'-type Galilean invariance in the primed inertial coordinate system plus the transformation from the primed to the usual inertial coordinate systems. The modified special relativity theory and the quantum mechanics theory together found a convergent and invariant quantum field theory.

The current Lorentz-invariant field theory, whether classical or quantum, has been suffering from the divergence difficulties for a long time. Indeed, the infinite self-energy of an electron in quantum electrodynamics was known as early as 1929 [1]. The origins of these difficulties lay deep within the conceptual foundations of the theory. Two foundation stones of the current quantum field theory are the special relativity theory and the quantum mechanics theory. Since it is the case that both classical field theory and quantum field theory are plagued by the divergence difficulties, the direction to get to the root of these difficulties seems to be in the special relativity theory. In this letter, we propose a modified special relativity theory and show how it, combined with the quantum mechanics theory, founds a convergent and invariant quantum field theory. The establishment of this quantum field theory does not demand departures from the concepts in the current quantum field theory such as local gauge symmetries, locality and local Lorentz invariance in the usual inertial coordinate systems, except for the radical change in our notion about local structures of gravity-free space and time.

Einstein published his special theory of relativity in 1905 [2]. He derived the Lorentz transformation between any two usual inertial coordinate systems, which is a kinematical background for the physical principle of the Lorentz invariance. The two fundamental postulates stated by Einstein as the basis for his theory are (i) the principle of relativity and (ii) the constancy of the one-way speed of light in all inertial frames of reference. Besides these two fundamental postulates, the special theory of relativity also uses another assumption. This other assumption concerns the Euclidean structure of gravity-free space and the homogeneity of gravity-free time in the usual inertial coordinate system $\{x^r,t\}$, r=1,2,3, $x^1=x$, $x^2=y$, $x^3=z$:

$$dX^2=\delta_{rs}dx^r dx^s, \ r,s=1,2,3, \qquad (1a)$$
$$dT^2=dt^2, \qquad (1b)$$

everywhere and every time.

Postulates (i) and (ii) together with the assumption Eqs.(1) yield the Lorentz transformation between any two usual inertial coordinate systems [2-4]. Though the assumption Eqs.(1) was not explicitly articulated, evidently having been considered self-evident, Einstein said in 1907: "Since the propagation velocity of light in empty space is c with respect to both reference systems, the two equations, $x_1^2+y_1^2+z_1^2-c^2t_1^2=0$ and $x_2^2+y_2^2+z_2^2-c^2t_2^2=0$, must be equivalent." [2]. Leaving aside a discussion of whether postulate (i) implies the linearity of transformation between any two usual inertial coordinate systems and the reciprocity of relative velocities between any two usual inertial coordinate systems, we know that the two equivalent equations, the linearity of transformation and the reciprocity of relative velocities lead to the Lorentz transformation. Some physicists explicitly articulated the assumption Eqs.(1) in their works on the topic. Pauli wrote: "This also implies the validity of Euclidean geometry and the homogeneous nature of space and time." [3]. Fock said: "The logical foundation of these methods is, in principle, the hypothesis that Euclidean geometry is applicable to real physical space together with further assumptions, viz. that rigid bodies exist and that light travels in straight lines." [4].

Introducing the four-dimensional usual inertial coordinate system $\{x^k\}$, k=1,2,3,4, $x^4$=ict, and the Minkowskian structure of gravity-free spacetime in this coordinate system,

$$d\Sigma^2=\delta_{ij}dx^i dx^j, \quad i,j=1,2,3,4, \tag{2}$$

Minkowski [5] showed in 1909 that the Lorentz transformation is just a rotation in this spacetime. He also showed how to use the four-dimensional tensor analysis for writing invariant physical laws with respect to the Lorentz transformation. The Minkowskian structure Eq.(2) is a four-dimensional version of the assumption Eqs.(1).

Conceptually, the principle of relativity means that there exists a class of equivalent inertial frames of reference, any one of which moves with a non-zero constant velocity relative to any other. Einstein wrote in his Autobiographical Notes: "in a given inertial frame of reference the coordinates mean the results of certain measurements with rigid (motionless) rods, a clock at rest relative to the inertial frame of reference defines a local time, and the local time at all points of space, indicated by synchronized clocks and taken together, give the time of this inertial frame of reference." [6]. As defined by Einstein, each of the equivalent inertial frames of reference is supplied with motionless, rigid unit rods of equal length and motionless, synchronized clocks of equal running rate. Then in each inertial frame of reference, an observer can employ his own motionless-rigid rods and motionless-synchronized clocks in the so-called "motionless-rigid rod and motionless-synchronized clock" measurement method to measure space and time intervals. By using this "motionless-rigid rod and motionless-synchronized clock" measurement method, the observer in each inertial frame of reference can set up his own usual inertial coordinate system $\{x^r,t\}$, r=1,2,3. Postulate (ii) means that the speed of light is measured to be the same constant c in every such usual inertial coordinate system. Recent null experiments searching for the anisotropy in the one-way speed of light support this postulate [7,8].

The "motionless-rigid rod and motionless-synchronized clock" measurement method is not the only one that each inertial frame of reference has. We imagine, for each inertial frame of reference, other measurement methods that are different from the "motionless-rigid rod and motionless-synchronized clock" measurement method. By taking these other measurement methods, an observer in each inertial frame of reference can set up other inertial coordinate systems, just as well as he can set up his usual inertial coordinate system by taking the "motionless-rigid rod and motionless-synchronized clock" measurement method. We call these other inertial coordinate systems the unusual inertial coordinate systems. Conventional believe in flatness of gravity-free space and time is natural. But question is, in which inertial coordinate system the gravity-free space and time directly display their flatness. The special theory of relativity recognizes the usual inertial coordinate system, as shown in the assumption Eqs.(1). Making a different choice, we take one of the unusual inertial coordinate systems, say $\{x'^r,t'\}$, r=1,2,3, the primed inertial coordinate system. We assume that gravity-free space and time possess the flat metric structures in the primed inertial coordinate system, and hence, the following generalized Finslerian structures in the usual inertial coordinate system [9,10],

$$dX^2=\delta_{rs}dx'^r dx'^s=g_{rs}(y)dx^r dx^s, \quad r,s=1,2,3, \tag{3a}$$
$$dT^2=dt'^2=g(y)dt^2, \tag{3b}$$
$$g_{rs}(y)=K^2(y)\delta_{rs}, \tag{3c}$$
$$g(y)=(1-y^2/c^2), \tag{3d}$$
$$K(y)=\frac{c}{2y}(1-y^2/c^2)^{1/2}\ln\frac{c+y}{c-y}, \tag{3e}$$

where $y=(y^s y^s)^{1/2}$, $y^s=dx^s/dt$, s=1,2,3.

We modify the special theory of relativity by combining the alternative assumption Eqs.(3), instead of the assumption Eqs.(1), with the two postulates (i) and (ii). If we define a new type of velocity, $y'^s=dx'^s/dt'$, s=1,2,3, in the primed inertial coordinate system and keep the well-defined usual (Newtonian) velocity $y^s=dx^s/dt$, s=1,2,3, in the usual inertial coordinate system, we find from the assumption Eqs.(3),

$$y'^s = \left[\frac{c}{2y} \ln \frac{c+y}{c-y}\right] y^s, \quad s=1,2,3, \tag{4}$$

and

$$y' = \frac{c}{2} \ln \frac{c+y}{c-y}, \tag{5}$$

where $y'=(y'^s y'^s)^{1/2}$ and $y=(y^s y^s)^{1/2}$, s=1,2,3. It is understood that two different measurement methods can be applied to a motion when it is observed in an inertial frame of reference, one being the "motionless-rigid rod and motionless-synchronized clock" measurement method, the other one being associated with the primed inertial coordinate system. As a result, two different velocities, usual velocity $y^s$ and primed velocity $y'^s$ (of the new type), are obtained. These two velocities are related by Eqs.(4) and (5). Velocities $y^s$ and $y'^s$ are two different versions of the motion obtained via two different measurement methods taken in the inertial frame of reference. Velocity $y'^s$, s=1,2,3, varies uniquely with $y^s$ and equals $y^s$ only when $y^s$ vanishes. The Galilean addition among primed velocities links up with the Einstein addition among usual velocities [11]. Actually, in the one-dimensional case, it is easily seen that

$$y'_2 = y'_1 - u' = (c/2)\ln[(c+y_1)/(c-y_1)] - (c/2)\ln[(c+u)/(c-u)]$$

and

$$y'_2 = (c/2)\ln[(c+y_2)/(c-y_2)]$$

imply

$$y_2 = (y_1 - u)/(1 - y_1 u/c^2).$$

In Eq.(5), as y goes to c, we get an infinite primed speed,

$$c' = \lim_{y \to c} \frac{c}{2} \ln \frac{c+y}{c-y}. \tag{6}$$

Speed c' is invariant in the primed inertial coordinate systems simply because of the invariance of speed c in the usual inertial coordinate systems. Speed c' is actually a new version of the light speed, its version in the primed inertial coordinate systems.

Let IFR1 and IFR2 be two inertial frames of reference, where IFR2 moves with a non-zero constant velocity relative to IFR1. IFR1 and IFR2 can use their own "motionless-rigid rod and motionless-synchronized clock" measurement methods and set up their own usual inertial coordinate systems $\{x^r_m, t_m\}$, m=1,2. They can also set up their own primed inertial coordinate systems $\{x'^r_m, t'_m\}$, m=1,2. Since the propagation velocity of light is c' in both $\{x'^r_1, t'_1\}$ and $\{x'^r_2, t'_2\}$, we have two equivalent equations,

$$\delta_{rs} dx'^r_1 dx'^s_1 - c'^2 (dt'_1)^2 = 0, \tag{7a}$$
$$\delta_{rs} dx'^r_2 dx'^s_2 - c'^2 (dt'_2)^2 = 0. \tag{7b}$$

Using Eqs.(3) with y=c, we have further two equivalent equations,

$$\delta_{rs} dx^r_1 dx^s_1 - c^2 (dt_1)^2 = 0, \tag{8a}$$
$$\delta_{rs} dx^r_2 dx^s_2 - c^2 (dt_2)^2 = 0, \tag{8b}$$

because $c^2 K^2(c) = c'^2 g(c)$, where $K(c) = \lim_{y \to c} K(y)$, $g(c) = \lim_{y \to c} g(y)$.

Two equivalent equations (7), the linearity of transformation between two $\{x'^r_m, t'_m\}$, the reciprocity of relative primed velocities between two $\{x'^r_m, t'_m\}$, and the flat structures of gravity-free space and time in two $\{x'^r_m, t'_m\}$ will lead to the c'-type Galilean transformation between two primed inertial coordinate systems $\{x'^r_m, t'_m\}$, under which speed c' is invariant. Two equivalent equations (8), the linearity of transformation between two $\{x^r_m, t_m\}$, and the reciprocity of relative usual velocities between two $\{x^r_m, t_m\}$ will lead to the localized Lorentz transformation between two usual inertial coordinate systems $\{x^r_m, t_m\}$, where the space and time differentials respectively take places of the space and time variables in the Lorentz transformation. The modified special relativity theory keeps the constancy of the one-way speed of light and the local Lorentz invariance in the usual inertial coordinate systems. Nevertheless, the modified special relativity theory involves two versions of the light speed, infinite speed c' in the primed inertial coordinate system and finite speed c in the usual inertial coordinate system. It involves the c'-type Galilean transformation between any two primed inertial coordinate systems and the localized Lorentz transformation between any two usual inertial coordinate systems.

The assumption Eqs.(3) has its four-dimensional version. Introducing the four-dimensional primed inertial coordinate system $\{x'^k\}$, k=1,2,3,4, $x'^4=ic't'$ and the four-dimensional usual inertial coordinate system $\{x^k\}$, k=1,2,3,4, $x^4=ict$, we may have

$$d\Sigma^2 = \delta_{ij}dx'^i dx'^j = g_{ij}(z)dx^i dx^j, \quad i,j=1,2,3,4, \tag{9a}$$

with

$$g_{11}(z)=g_{22}(z)=g_{33}(z)=K^2(z),\ g_{44}(z)=G^2(z),\ g_{ij}(z)=0 \text{ for } i\neq j, \tag{9b}$$

$$K(z)=\frac{c}{2z}\ln\frac{\sqrt{c^2+z^2}+z}{\sqrt{c^2+z^2}-z}, \tag{9c}$$

$$G(z)=\frac{c}{2\sqrt{c^2+z^2}}\lim_{z\to+\infty}\ln\frac{\sqrt{c^2+z^2}+z}{\sqrt{c^2+z^2}-z}, \tag{9d}$$

where $z=(z^s z^s)^{1/2}$, s=1,2,3, and

$$z^r \equiv dx^r/dt' = y^r/(1-y^2/c^2)^{1/2},\ r=1,2,3, \tag{10a}$$
$$z^4 \equiv dx^4/dt' = ic/(1-y^2/c^2)^{1/2}. \tag{10b}$$

Metric tensors $g_{rs}(y)$ and $g(y)$ in Eqs.(3) or metric tensor $g_{ij}(z)$ in Eqs.(9) depend or depends only on directional variables $y^s$ or $z^s$, s=1,2,3. The geometry specified by metric tensors $g_{rs}(y)$ and $g(y)$ or metric tensor $g_{ij}(z)$ can be studied and described through the generalized Finsler geometry [12].

Finsler geometry is a kind of generalization of Riemann geometry, while generalized Finsler geometry is a kind of generalization of Finsler geometry. In generalized Finsler geometry, distance ds between two neighboring points $x^k$ and $x^k+dx^k$, k=1,2,---,n is defined by

$$ds^2 = g_{ij}(x^k, dx^k)dx^i dx^j, \tag{11}$$

where metric tensor $g_{ij}(x^k,dx^k)$ depends on directional variables $dx^k$ as well as coordinate variables $x^k$ and satisfies

$$g_{ij}(x^k,dx^k) = g_{ji}(x^k,dx^k), \tag{12a}$$
$$g_{ij}(x^k,\lambda dx^k) = g_{ij}(x^k,dx^k) \text{ for } \lambda>0. \tag{12b}$$
$$\det[g_{ij}(x^k,dx^k)] \neq 0, \text{ and} \tag{12c}$$
$$g_{ij}(x^k,dx^k)dx^i dx^j > 0,\ i,j=1,2,---,n, \text{ for non-zero vector } dx^i. \tag{12d}$$

Like Finsler geometry, the generalized Finsler geometry can be endowed with the Cartan connection [12]. In our case that metric tensor $g_{ij}(z)$, i,j=1,2,3,4, depends only on four-dimensional directional variables $z^k$, k=1,2,3,4, according to the Cartan connection, we have

$$Dz^k = dz^k,\ k=1,2,3,4, \tag{13}$$
$$\nabla_h = \partial_h,\ h=1,2,3,4, \tag{14}$$

where $Dz^k$ is the absolute differential of $z^k$ and $\nabla_h$ is the covariant partial derivative of the first kind [10].

Since it is the "motionless-rigid rod and motionless-synchronized clock" measurement method that we use in our experiments, the physical principle in the modified special relativity theory is: The c'-type Galilean invariance in the primed inertial coordinate system plus the transformation from the primed inertial coordinate system to the usual inertial coordinate system. In the primed inertial coordinate system, we write all physical laws in the c'-type Galilean-invariant form, we do calculations in the c'-type Galilean-invariant manner, and we finally transform all results from the primed inertial coordinate system to the usual inertial coordinate system and compare them to experimental facts in the usual inertial coordinate system. The transformation from the primed to the usual inertial coordinate systems is an important part of the new physical principle. If two axes $x'^r$ and $x^r$, where r runs 1,2,3, are set to have the same direction and the same origin, the transformation is

$$dx'^r = K(y)dx^r,\ r=1,2,3, \tag{15a}$$

and

$$dt' = (1-y^2/c^2)^{1/2}dt \tag{15b}$$

or

$$dx'^4 = G(y)dx^4,\ G(y)=\frac{c'}{c}(1-y^2/c^2)^{1/2}. \tag{15c}$$

This transformation links the Galilean transformation between any two primed inertial coordinate systems up with the localized Lorentz transformation between two corresponding usual inertial coordinate systems [10].

The light speed that owns the constancy in all inertial frames of reference has two versions: infinite speed c' in the primed inertial coordinate system and finite speed c in the usual inertial coordinate system. We deal with this constant in such a way that it has the value of c' in the primed inertial coordinate system and of c in the usual inertial coordinate system.

Now we apply this principle to reform of mechanics and field theory. The c'-type Galilean-invariant motion equation in the primed inertial coordinate system is

$$F'^k = m_0 dz'^k/dt', \quad k=1,2,3,4, \tag{16}$$

where $m_0$ is the mass of an object, $F'^k$ is the four-dimensional force, $z'^k = dx'^k/dt'$, k=1,2,3,4, and the locally Lorentz-invariant motion equation in the usual inertial coordinate system is

$$F^k = m_0 Dz^k/dt', \quad k=1,2,3,4. \tag{17}$$

Using Eq.(13), we obtain

$$F^k = m_0 dz^k/dt', \quad k=1,2,3,4, \tag{18}$$

as the motion equation in the usual inertial coordinate system.

In the primed inertial coordinate system, the c'-type Galilean-invariant four-dimensional energy-momentum vector of a moving particle with rest mass $m_0$ is

$$p'^k = m_0 z'^k = (m_0 y'^1, m_0 y'^2, m_0 y'^3, im_0 c'), \quad k=1,2,3,4. \tag{19}$$

Acting the transformation from the primed to the usual inertial coordinate systems on it, we find

$$p'^k \to p^k = (m_0 y'^r / K(y), im_0 c' / G(y)) = (m_0 y^r / [1-y^2/c^2]^{1/2}, im_0 c / [1-y^2/c^2]^{1/2})$$
$$= m_0 z^k, \quad k=1,2,3,4, \quad r=1,2,3, \tag{20}$$

as the four-dimensional energy-momentum vector in the usual inertial coordinate system. Immediately calculating the trace of tensor $p^i p^j$, i,j=1,2,3,4, yields

$$Sp(p^i p^j) = -m_0^2 c^2, \quad y<c, \tag{21}$$
$$0, \quad y=c.$$

Its familiar form is

$$E^2 = c^2 \overline{p} \cdot \overline{p} + m_0^2 c^4,$$

where $\overline{p} = (p^1, p^2, p^3)$, $E = -icp^4$, and E is the energy of the moving particle.

The change of assumption from Eqs.(1) to Eqs.(3) has no effect on the validity of relativistic mechanics in the usual inertial coordinate system. However, the equations regarding the invariant of $\delta_{ij} p'^i p'^j = g_{ij}(z) p^i p^j$ alter,

$$\delta_{ij} p'^i p'^j = -m_0^2 c'^2, \quad y'<c', \tag{22}$$
$$0, \quad y'=c',$$

in the primed inertial coordinate system and

$$g_{ij}(z) p^i p^j = -m_0^2 c^2, \quad y<c, \tag{23}$$
$$0, \quad y=c,$$

in the usual inertial coordinate system.

Owing to Eqs.(14), we have the following interaction-free field equations

$[g^{ij}(z) \partial_i \partial_j - æ^2]\varphi = 0$ for massive real scale field $\varphi$,

$\delta^{ij} \partial_i \partial_j A_k = 0$, k=1,2,3,4, for massless vector field $A_k$,

$[g^{ij}(z) \partial_i \partial_j - æ^2] A_k = 0$, k=1,2,3,4, for massive vector field $A_k$,

$[g^{ij}(z) \gamma_i \partial_j - æ]\psi = 0$ and

$\overline{\psi} [g^{ij}(z) \gamma_i \partial_j - æ] = 0$ for massive even-spinor field $\psi$,

in the usual inertial coordinate system, where $æ = m_0 c/\hbar$, $\gamma_s$, s=1,2,3, and $\gamma_4$ are the Dirac matrices in the usual inertial coordinate system, $\{\gamma_i, \gamma_j\} = 2g_{ij}(z)$.

These interaction-free field equations keep the de Broglie wave solution,

$$\approx \exp[\frac{i}{\hbar} p_i x^i], \quad i=1,2,3,4. \tag{24}$$

In the primed inertial coordinate system, the canonically conjugate variables to field variables $\varphi'_k$ (vector or spinor) are

$$\pi'^k = \partial L'/\partial(\partial \varphi'_k/\partial t'), \quad (25)$$

while those to field variables $\varphi_k$ in the usual inertial coordinate system are

$$\pi^k = \partial L/\partial(\partial \varphi_k/\partial t'). \quad (26)$$

The Noether theorem is

$$\frac{d}{dt'}\left[\frac{1}{ic'}\int f'^4 d\mathbf{x}'\right] = 0 \quad (27)$$

in the primed inertial coordinate system and

$$\frac{d}{dt'}\left[\frac{1}{ic}\int f^4 d\mathbf{x}'\right] = 0 \quad (28)$$

in the usual inertial coordinate system, where $d\mathbf{x}' = dx'^1 dx'^2 dx'^3$, $f'^4$ and $f^4$ are respectively the fourth component of vectors

$$f'^i = \{L'\delta^i_j - [\partial L'/\partial(\partial \varphi'_k/\partial x'^i)][\partial \varphi'_k/\partial x'^j]\}\delta x'^j + [\partial L'/\partial(\partial \varphi'_k/\partial x'^i)]\delta \varphi'_k, \quad i,j,k=1,2,3,4,$$

and

$$f^i = \{L g^i_j(z) - [\partial L/\partial(\partial \varphi_k/\partial x^i)][\partial \varphi_k/\partial x^j]\}\delta x^j + [\partial L/\partial(\partial \varphi_k/\partial x^i)]\delta \varphi_k, \quad i,j,k=1,2,3,4.$$

In particular, the conserved field energy and momentum in the usual inertial coordinate system is

$$P_j = \frac{1}{ic}\int T^4_j d\mathbf{x}', \quad j=1,2,3,4,$$

where

$$T^4_j = L g^4_j(z) - [\partial L/\partial(\partial \varphi_k/\partial x^4)][\partial \varphi_k/\partial x^j].$$

The gauging procedure adopted in the current field theory to make a field system locally gauge-invariant with respect to a certain gauge group is still effective in the primed inertial coordinate system. As we perform the transformation from the primed to the usual inertial coordinate systems, we can find the version of this procedure in the usual inertial coordinate system. For example, in the primed inertial coordinate system, the U(1) gauge transformation and the U(1) gauge-invariant interaction are

$$\psi'_\alpha \to \psi'_\alpha \exp[i\lambda'(x')] \text{ and } A'_h \to A'_h - \frac{1}{e}\partial'_h \lambda'(x'), \quad (29a)$$

$$L'_{int} = ie\overline{\psi}'\delta^{ij}\gamma_i A'_j \psi', \quad (29b)$$

where $\lambda'(x')$ is a real scale function. Consequently, in the usual inertial coordinate system, they are

$$\psi_\alpha \to \psi_\alpha \exp[i\lambda(x)] \text{ and } A_h \to A_h - \frac{1}{e}\partial_h \lambda(x), \quad (30a)$$

$$L_{int} = ie\overline{\psi} g^{ij}(z)\gamma_i A_j \psi, \quad (30b)$$

where Eqs.(14) are used.

It is the most remarkable feature of the field theory in the framework of the modified special relativity theory that the concept of particle size has its own room. All particles properly display themselves in the primed inertial coordinate system. Particle size defined in the primed inertial coordinate system is an invariant quantity. It can be quite involved in our invariant calculations. It should be noted, however, that the concept of particle size in the primed inertial coordinate system is different from that we tried but failed to introduce in the current Lorentz-invariant field theory. We shall discuss it in detail elsewhere.

In the primed inertial coordinate system, any field system can be quantized by use of the canonical quantization method. As instantaneity is a covariant concept, the equal-time commutation or anti-commutation relations are reasonable. As canonically conjugate variables are contravariant to original field variables, the commutation or anti-commutation relations are also of tensor equations,

$$[\varphi'_\sigma(\mathbf{x}',t'), \pi'^\rho(\mathbf{x}'+\delta\mathbf{x}',t')]_\mp = i\hbar\, \delta_\sigma^{\;\rho}\delta^3(\delta\mathbf{x}'). \tag{31}$$

As the light speed c' is infinite, the essentially instantaneous quantum connection is acceptable. As the primed time differential dt' is invariant, the time-ordered product in the perturbation expansion of S-matrix is definite. In the primed inertial coordinate system, the state vector equation of a quantized field system is

$$i\hbar \frac{d}{dt'}\Phi' = H'\Phi', \tag{32}$$

while its operator evolution satisfies

$$\frac{d}{dt'}\Omega' = \frac{i}{\hbar}[H', \Omega'], \tag{33}$$

where H' is the Hamiltonian of the system, Φ' a state vector and Ω' an operator.

How is a quantized field system transformed under the transformation from the primed to the usual inertial coordinates systems? If we denote the operator of this quantized field system in the primed inertial coordinate system by Ω' and the state vector by Φ', and denoting those of the transformed quantized field system in the usual inertial coordinate system by Ω and Φ, how are Ω' and Φ' related to Ω and Φ? A transformation law has been established to answer this question [10]. It says: any quantized field system will undergo a unitary transformation under the transformation from the primed inertial coordinate system to the usual inertial coordinate systems. In general, for a quantized field system, we can find a unitary transformation U such that

Ω'=UΩU$^{-1}$, Φ'=UΦ.

The transformation law allows us immediately to write, from Eqs.(32) and (33),

$$i\hbar \frac{d}{dt'}\Phi = H\Phi \tag{34}$$

and

$$\frac{d}{dt'}\Omega = \frac{i}{\hbar}[H, \Omega] \tag{35}$$

as the state vector equation and operator equation in the usual inertial coordinate system, and

$$[\varphi_\sigma(\mathbf{x},t), \pi^\rho(\mathbf{x}+\delta\mathbf{x},t)]_\mp = i\hbar\, \frac{1}{\sqrt{g}}\delta_\sigma^{\;\rho}\delta^3(\delta\mathbf{x}), \tag{36}$$

where g=det[$g_{rs}$(z)], r,s=1,2,3, as the commutation or anti-commutation relations in the usual inertial coordinate system.

The divergence difficulties in the current Lorentz-invariant quantum field theory have been already ascribed to the model of point particle. But all attempts to assign a finite size to a particle failed. According to the special relativity theory, particle size is not a covariant concept. It has been unknown how to carry out the Lorentz-invariant calculations based on such a size. The Lorentz invariance recognizes the light speed c constituting a limitation for transport of matter or energy and transmission of information or causal connection. The law of causality rejects any instantaneous processes between two events at distinct space points. It is very hard to explain how a sized particle as a whole is set in motion when a force acts on it at its edge. It is very hard to explain how the quantum connection specified by the quantum mechanics theory can be instantaneous. Also, as pointed out by Poincare [13], the field energy and momentum of a sized electron, in the framework of the special relativity theory, do not have correct transformation properties, unlike its mechanical energy and momentum, though they are finite. That will result in different values when we compute some quantities, f.g. the total mass of the sized electron.

All of these alter in the modified special relativity theory. In the primed inertial coordinate system, the concept of particle size is a proper concept, it is c'-type Galilean-invariant. The light speed constituting a limitation for the matter or energy transport and the information or causal connection transmission in the primed inertial coordinate system is c', infinite. The field energy and momentum of a sized particle form a vector as well as its mechanical energy and momentum. In the primed inertial coordinate system, all particles properly exhibit their own size. As a matter of course, we have a convergent and c'-type Galilean-invariant quantum field theory for any quantized field system in the primed inertial coordinate system. When we perform the transformation from the primed to the usual inertial coordinate systems, the unitary transformation does not change the eigenvalue spectrums of operators of this quantized field system, the expectation values of its observations, and its operator and state vector equations. We hence have a convergent and invariant quantum field theory for the quantized field system in the usual inertial coordinate system. Since it is the localized Lorentz transformation that stands between any two usual inertial coordinate systems, we indeed have a convergent and locally Lorentz-invariant quantum field theory for the quantized field system in the usual inertial coordinate system.

The modified special relativity theory and the quantum mechanics theory together found a convergent and invariant quantum field theory.

ACKNOWLEDGMENT
The author greatly appreciates the teachings of Prof. Wo-Te Shen. The author thanks Prof. Mark Y.-J. Mott and Dr. Allen E. Baumann for helpful suggestions.

REFERENCES
[1] W. Heisenberg and W. Pauli, Z. Phys., 56, 1 (1929); 59, 168 (1930); I. Waller, Z. Phys., 62, 673 (1930); J. R. Oppenheimer, Phys. Rev., 35, 461 (1930); W. Heisenberg, Z. Phys., 90, 209 (1934); P. A. M. Dirac, Pro. Camb. Phil. Soc., 30, 150 (1934)
[2] A. Einstein, Ann. Physik, 17, 891 (1905); A. Einstein, Jarbuch der Radioaktivitat und Elektronik, 4, 411 (1907), reprinted in The Collected Papers of A. Einstein, vol.2, 252, Princeton University Press, Princeton, NJ (1989)
[3] A. Einstein, H. A. Lorentz, H. Minkowski and H. Weyl, The Principle of Relativity, collected papers with notes by A. Sommerfield, Dover, New York (1952); A. S. Eddington, The Mathematical Theory of Relativity, Cambridge University Press, Cambridge (1952); C. Moller, The Theory of Relativity, Oxford and New York (1952); W. Pauli, Theory of Relativity, Pergamom Press Ltd., New York (1958), English translator: G. Field
[4] V. Fock, The Theory of SpaceTime and Gravitation, Pergamon Press, New York (1959)
[5] H. Minkowski, Phys. Z., 10, 104 (1909)
[6] A. Einstein, Autobiographical Notes, in: A. Einstein: Philosopheo-Scientist, ed. P. A. Schipp, 3rd edition, Tudor, New York (1970)
[7] R. M. Barnett et al, Rev. Mod. Phys., 68, 611 (1996), p655
[8] K. C. Turner and H. A. Hill, Phys. Rev., 134, B252 (1964); D. C. Champeney, G. R. Isaak and A. M. Khan, Phys. Lett., 7, 241 (1963); E. Riis et al, Phys. Rev. Lett., 60, 81 (1988); T. P. Krisher et al, Phys. Rev., D45, 731 (1990); R. W. McGowan et al, Phys. Rev. Lett., 70, 251 (1993); J. Bailey et al, Nature (London), 268, 301 (1977); M. Kaivola et al, Phys. Rev. Lett., 54, 255 (1985); J. D. Prestage et al, Phys. Rev. Lett., 54, 2387(1985)
[9] Jian-Miin Liu, Galilean Electrodynamics (Massachusetts, USA), 8, 43 (1997); 9, 73 (1998); On local structures of gravity-free space and time, e-printed in physics/9901001 (1999)
[10] Jian-Miin Liu, Modification of special relativity and formulation of convergent and invariant quantum field theory, e-printed in hep-th/9805004 (1998)
[11] Jian-Miin Liu, Velocity-space in the modified special relativity theory, to be published.
[12] P. Finsler, Uber Kurven und Flachen in Allgemeinen Raumen, Dissertation, Gottingen 1918, Birkhauser Verlag, Basel (1951); E. Cartan, Les Espaces de Finsler, Actualites 79, Hermann, Paris (1934); H. Rund, The Differential Geometry of Finsler Spaces, Springer-Verlag, Berlin (1959); G. S. Asanov, Finsler Geometry, Relativity and Gauge Theories, D. Reidel Publishing Company, Dordrecht (1985); Jian-Miin Liu, The generalized Finsler geometry, to be published
[13] H. Poincare, Comtes Rendus (Paris), 40, 1504 (1905); La Mechanique Nouvelle, Gauthier-Villars (Paris, 1924)